\documentclass[prc,twocolumn,showpacs,amssymb]{revtex4}

\usepackage{graphics}
\usepackage{epsfig}

\usepackage{ulem}
\usepackage{amsmath}
\usepackage {float}
\begin{document}

\title{Two-neutron transfer in Sn isotopes beyond the $N=82$ shell closure} 

\author{B. F. Bayman$^{1}$, A. Covello$^{2}$, A. Gargano$^{3}$,
P. Guazzoni$^{4}$, and L. Zetta$^{4}$} 
\affiliation{$^{1}$School of Physics and Astronomy, University of Minnesota, Minneapolis, Minnesota 55455, USA \\
$^{2}$Dipartimento di Fisica, Universit\`a
di Napoli Federico II,
Complesso Universitario di Monte S. Angelo, Via Cintia, I-80126 Napoli,
Italy \\
$^{3}$Istituto Nazionale di Fisica Nucleare, 
Complesso Universitario di Monte S. Angelo, Via Cintia,  I-80126 Napoli,
Italy\\
$^{4}$Dipartimento di Fisica dell'Universit\`a, and Istituto Nazionale di Fisica Nucleare, Via Celoria 16, I-20133 Milano, Italy}

\date{\today}

\begin{abstract}

We have performed microscopic distorted-wave Born approximation (DWBA) calculations of differential cross sections for the two reactions $^{136}$Sn($p,t$)Sn$^{134}$ and $^{134}$Sn($t,p$)Sn$^{136}$, which are within reach of near-future experiments with radioactive ion beams.
We have described the initial and final nuclear states in terms of the shell model, 
employing a realistic low-momentum two-body effective interaction derived from the CD-Bonn nucleon-nucleon potential that has already proved quite successful in describing the available low-energy energy spectrum of $^{134}$Sn.
We discuss the main features of the predicted cross sections for the population of the low-lying yrast states in the two nuclei considered.

\end{abstract}    


\maketitle

\section{Introduction}
Two-neutron transfer reactions have long been recognized as a most valuable tool to gain information on nuclear structure near closed shells. In fact, the first study with ($p,t$) reactions on all the stable even Sn isotopes dates back to 1970 \cite{Fleming70}. Some thirty years later a systematic study of these isotopes via ($p,t)$ reactions was undertaken in high resolution experiments at the Munich HVEC MP Tandem, which led to identification of many new low-spin excited levels. The results for the reactions $^{112,114,116,118,120,122,124}$Sn($p,t$)Sn$^{110,112,114,116,118,120,122}$ were reported in various papers 
\cite{Guazzoni06,Guazzoni12,Guazzoni04,Guazzoni11,Guazzoni08,Guazzoni99},  where they were also compared with predictions of shell-model calculations with realistic effective interactions. 

In all of these papers, except Refs. \cite{Guazzoni99,Guazzoni04}, a microscopic DWBA calculation of differential cross sections was performed, where we used two-neutron spectroscopic amplitudes obtained from shell-model wave functions for the initial and final nuclear states calculated in a seniority space including states up to seniority $v=4$. The agreement of the calculated spectra with the experimental ones was on the whole remarkably good. As regards the cross section angular distributions  for the excitation of the lowest yrast states of the final nucleus,   
they are well reproduced in general, the most significant disagreement between theory and experiment occurring for the ($p,t$) reaction leading to the mid-shell nucleus $^{116}$Sn. 

To summarize, we may say that a main outcome of our detailed study was that it provided further evidence of the key role played by the shell model in understanding the structure of the tin isotopic chain.  The latter, however, extends well beyond the last stable isotope, $^{124}$Sn, spectroscopic information being now available beyond $^{132}$Sn. More precisely, four excited states and two B(E2) transition rates are known in $^{134}$Sn, while only one $2^+$ state has been very recently identified at 682 keV excitation energy \cite{Wang14} in $^{136}$Sn.
This excitation energy is slightly smaller than that of the $2^+$ state in $^{134}$Sn (726 keV),
making it the lowest first excited $2^+$ level observed in a semi-magic even-even nucleus over the whole chart of nuclides. The spectrum and B(E2) transition rates in $^{134}$Sn were successfully reproduced by  shell-model calculations  \cite{Covello11},
employing a realistic effective interaction derived from the CD-Bonn nucleon-nucleon ($NN$) potential \cite{Machleidt01}.  The spectrum of $^{136}$Sn was predicted in \cite{Covello11} and the calculated energy of the 2$^+$ state turns out to be in very good agreement with the recent observed value. The low position of this 2$^+$ state as well as of that in $^{134}$Sn were explained as a manifestation of a weak neutron pairing for  $N>82$, which was traced to a reduction of the effects of the one particle-one hole core excitations, as discussed in detail in Ref. \cite{Covello13}.

The advent of radioactive ion beams (RIBs) provides the opportunity to study nuclei far from stability with transfer reactions performed in inverse kinematics. This makes it possible to investigate Sn isotopes with increasing nucleon number via two-neutron transfer reactions. In this context, particularly challenging is the perspective offered by the second-generation RIB facilities to go beyond the shell closure at $N=82$.  

Based on our studies of the Sn isotopes, we think it would be useful  to make predictions of  
cross sections for the two reactions $^{136}$Sn($p,t$)$^{134}$Sn and $^{134}$Sn($t,p$)$^{136}$Sn.
This is a subject of great current interest, especially with regard to the latter reaction, which is likely to be performed in the near future employing a $^{134}$Sn RIB impinging on a tritium gas target. Recently, the two-neutron transfer mode in neutron-rich Sn isotopes beyond 
$N=82$ has been studied \cite{Pllumbi11,Shimoyama11,Shimoyama13} with emphasis put on the character of pairing correlations. In these studies, microscopic nuclear structure calculations have been performed within the Hartree-Fock-Bogoliubov (HFB) plus quasiparticle random phase approximation (QRPA) approach. This has clearly increased our motivation for the present study,
where the structures of $^{134}$Sn and $^{136}$Sn are described in terms of the shell model with  a realistic effective interaction.

\section{Outline of calculations}

Our calculation of the differential cross sections for the ($p,t$) and ($t,p$)
reactions uses target and residual nucleus wave functions obtained from a  shell-model calculation in which $^{132}$Sn is assumed as  a closed core with the valence neutrons occupying the six levels $0h_{9/2}$, $1f_{7/2}$, $1f_{5/2}$, $2p_{3/2}$, $2p_{1/2}$,  and $0i_{13/2}$ of the $82-126$ shell. The adoped Hamiltonian is the same we employed in our recent shell-model studies of neutron-rich nuclei beyond $^{132}$Sn \cite{Covello11,Coraggio13,Coraggio13a,Coraggio13b}. 

For the sake of completeness, we  give here a few details of this Hamiltonian.
The single-neutron energies have been taken from the experimental spectrum of $^{133}$Sn \cite{NNDC}, with the exception of the $0i_{13/2}$ level  which was determined from the position of the 10$^{+}$ state at 2.423 MeV in $^{134}$Sb. The adopted values, relative to the $1f_{7/2}$ level, are (in MeV): $\epsilon_{2p_{3/2}}=0.854$, $\epsilon_{2p_{1/2}}=1.363$, 
$\epsilon_{0h_{9/2}}=1.561$, $\epsilon_{1f_{5/2}}=2.005$, and $\epsilon_{0i_{13/2}}=2.690$. Note that  the $2p_{1/2}$ energy is that measured in the  experiment of Ref. \cite{Jones10}, where  the spectroscopic amplitudes of the  $1f_{7/2}$, $2p_{3/2}$, $2p_{1/2}$, $1f_{5/2}$ levels  were extracted. The obtained values  provide a clear confirmation of the single-particle nature of these four levels.  

 As mentioned in the Introduction, the two-body effective interaction
between the valence neutrons is derived from the CD-Bonn $NN$ potential, which is renormalized by means of the $V_{\rm low-k}$ approach \cite{Bogner02} with a cutoff momentum $\Lambda$ = 2.2 
fm$^{-1}$. The obtained low-momentum potential is then used to 
derive the two-body effective interaction  $V_{\rm eff}$ within the framework of the ${\hat Q}$-box folded diagram expansion \cite{Coraggio09,Coraggio12}, including diagrams up to second order in $V_{\rm low-k}$.  These diagrams are computed within the harmonic-oscillator basis using intermediate states composed of all possible hole states and particle states restricted to the five proton and neutron shells above the Fermi surface. The oscillator parameter is 7.88 MeV, as obtained from the expression  $\hbar \omega = 45 A^{-1/3} -25 A^{-2/3}$ with $A=132$.
The shell-model calculations have been performed by the NUSHELL code \cite{NuShell}.

Theories of direct ($p,t$) and ($t,p$) reactions between multi-nucleon states have been described elsewhere \cite{Yoshida62,Glendenning65,Bayman68,Guazzoni08}. The basic assumption is that the interaction responsible for the process involves only the degrees of freedom of the proton and the transferred neutrons. The remaining target nucleons contribute only via the shell-model potential, which governs the motion of the bound neutron states, and the optical potentials, which govern the scattering states of the proton and triton. This assumption allows the differential cross-section for the transition between the ${\rm(p,I_1)}$ and ${\rm(t,I_2)}$ systems to be expressed in the form

\begin{widetext}
\begin{equation}
\frac{d\sigma}{d\Omega}\left(\theta\right)=\sum_L\left|\sum_{n_1,\ell_1,j_1,n_2,\ell_2,j_2} S\left(n_1,\ell_1,j_1,n_2,\ell_2,j_2;L;I_1,I_2\right)\times f\left(n_1,\ell_1,j_1,n_2,\ell_2,j_2;L;\theta \right)\right|^2.
\end{equation}
\end{widetext}

The spectroscopic amplitude $S\left(n_1,\ell_1,j_1,n_2,\ell_2,j_2;L;I_1,I_2\right)$ contains the nuclear structure information. It is defined as the matrix element of a vector-coupled pair of neutron creation operators between initial and final nuclear states:

\begin{widetext}

$$
S\left(n_1,\ell_1,j_1,n_2,\ell_2,j_2;L;I_1,I_2\right)\equiv <\Psi^{I_1}_{M_1}|\left[~\frac{\left[(a^{n_1,\ell_1,j_1})^{+}(a^{n_2,\ell_2,j_2})^{+}\right]^L}{\sqrt{1+\delta_{n_1,n_2}\delta_{\ell_1,\ell_2}\delta_{j_1,j_2}}}~\Psi^{I_2}~\right]^{I_1}_{M_1}~>
$$
\end{widetext}

Here the $(a^{n_1,\ell_1,j_1})^{+}$ and $(a^{n_2,\ell_2,j_2})^{+}$ represent neutron creation operators, and square brackets symbolize vector coupling. The spectroscopic amplitudes can be calculated once the shell-model wave functions $\Psi^{I_1}_{M_1},~\Psi^{I_2}_{M_2}$ are known. The reaction amplitudes $f\left(n_1,\ell_1,j_1,n_2,\ell_2,j_2;L;\theta\right)$ are calculated by a DWBA program. We used TWOFNR, written by M. Igarashi 
\cite{Igarashi77}, modified so as to allow the volume and surface contributions to the imaginary potentials to have different geometry. The reaction amplitudes depend upon the reaction kinematics, the optical potentials, and also on the details of the assumed reaction mechanism.
We have chosen optical potential parameters obtained by global fits to proton \cite{Becchetti69} and triton \cite{Li07} elastic scattering on medium-weight nuclei, at a range of energies.

\begin{figure}
\begin{center}
\includegraphics [scale=0.25,angle=0] {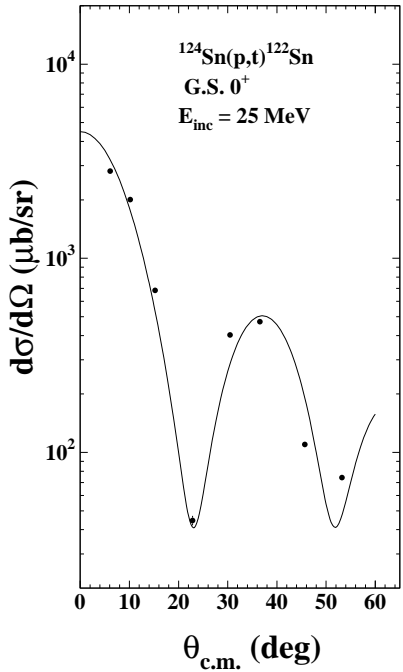}
\end{center}
\vspace{-0.8truecm}
\caption{\label{fig1}  Comparison between experimental \cite{Guazzoni11} and calculated differential cross-sections for the $^{124}$Sn(p,t)$^{122}$Sn ground-state transition. The line represents the results of the microscopic calculation (in $\mu{\rm b}/{\rm sr}$) as a function  of the center-of-mass angle (in degrees).}
\end{figure}

In our work, we have used the simple approximation that the interaction between the proton and the neutrons is a function only of the distance between the proton and the dineutron mass center. This implies that the relative motion of the two neutrons does not change during the transfer process, and so the reaction favors initial and final nuclear states which differ by a neutron pair with the same relative motion as the two neutrons in the triton. Since this is mostly a state with zero two-neutron spin and relative angular momentum, this leads to dominant transitions between states with strong pairing correlations. This has been an observed feature of ($p$,$t$) and ($t$,$p$) reactions once it became possible to resolve individual final levels 
\cite{Brink05}.

A major disadvantage of the proton-dineutron approximation is that it is difficult to normalize it, to yield absolute differential cross-sections. We can only calculate angular shapes of the differential cross-sections, and the relative strengths of transitions between specified nuclear states. Figure 1 shows a comparison with experimental data \cite{Guazzoni11} of our calculated differential cross-section for the $^{124}$Sn($p,t$)$^{122}$Sn ground-state transition at 25 MeV incident energy. Evidently the simple direct-interaction theory is able to give a good account of the shape of the angular distribution. We have normalized our calculation to produce a good visual fit to the magnitude of the data. We use this same normalization factor in the remainder of this paper, thereby allowing ourselves to express predicted differential cross-sections in $\mu$b/sr. Thus we are assuming that the normalization factor is the same, not only between levels of the same nucleus, but between the different tin isotopes. From the direct-reaction viewpoint, where the interaction involves only the proton and the transferred neutrons, this assumption of a constant normalization factor seems reasonable. 

In recent years, more precise calculations have been done in which the proton interacts separately with the individual neutrons \cite{Potel13}, and these calculations have succeeded in calculating absolute cross-sections, still within the picture of a direct reaction. 

The most significant feature of Eq. (1) is that the differential cross-section involves a {\it coherent} sum of contributions from the different $n_1,\ell_1,j_1;n_2,\ell_2,j_2$ pair transfers. This leads to strong cross-section variation from level to level, depending upon whether the different $n_1,\ell_1,j_1;n_2,\ell_2,j_2$ pairs contribute constructively or destructively. Thus two-neutron transfer reactions provide a very stringent test of calculated shell-model wave functions.

\section{Results and discussion}

As already mentioned in the Introduction, our shell-model study of Sn isotopes beyond $N=82$ [9,10] has led to remarkably good agreement between the calculated energy levels and the few experimental data available for  $^{134}$Sn and $^{136}$Sn. For the sake of completeness we report here in Fig. 2 the calculated excitation energies of the yrast $2^+$, $4^+$, and $6^+$ levels, as well as the second $0^+$ states for these two nuclei. These  will be the subject of the following discussion concerning the two-particle transfers. Also shown in Fig. 2 are the available experimental energies. We see that they are very well reproduced by the theory, the largest discrepancy being about 100 keV for the $6^+$ state in $^{134}$Sn.  

\vspace{0.5truecm}
\begin{figure} [H]
\begin{center}
\includegraphics [scale=0.50,angle=0] {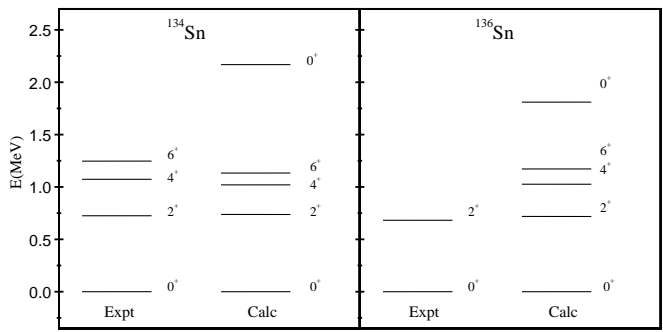}
\end{center}
\caption{\label{fig2} Calculated excitation energies of $^{134}$Sn and $^{136}$Sn compared with the available experimental ones.}
\end{figure}

\begin{figure} [H]
\begin{center}
\includegraphics [scale=0.30,angle=0] {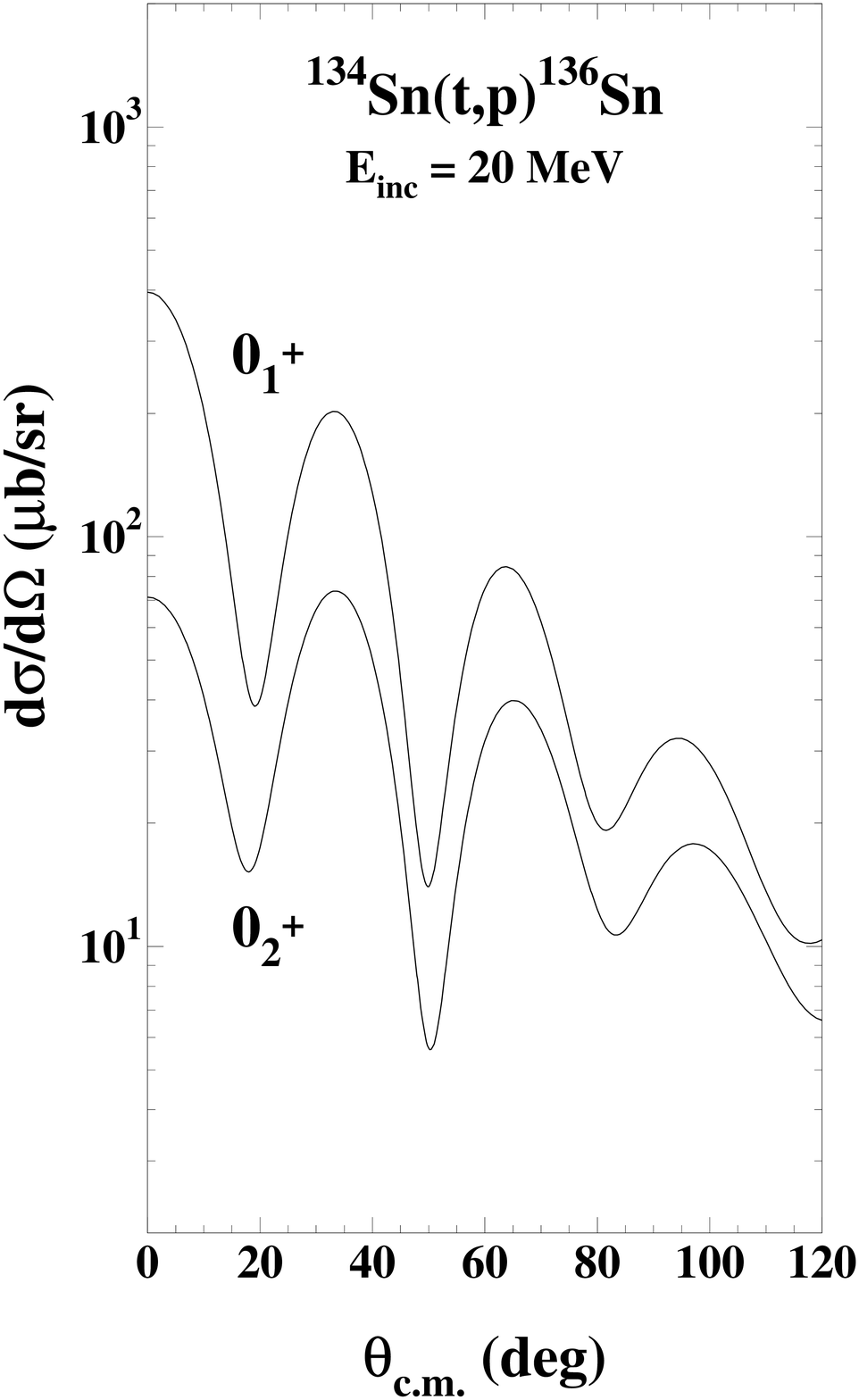}
\end{center}
\vspace{-1.2truecm}
\caption{\label{fig3} Calculated differential ($t$,$p$) cross sections  for the two lowest $0^+$ states in $^{136}$Sn at 20 MeV incident energy.}
\end{figure}

We start by focusing on the stripping and pick-up processes for  $^{134}$Sn and $^{136}$Sn leading to the population of the ground and the first excited $0^+$ states  of the final nucleus. This choice is motivated by the possibility to have a comparison with the results of Refs. \cite{Pllumbi11,Shimoyama11}. In these studies only transfers of nucleonic pairs to low-lying $0^+$ states  were investigated, with the aim of exploring the properties of the pairing force for neutron-rich nuclei. In this case, changes in the two-nucleon transfers are expected which may give important information on the nature of the pairing correlations. However, no experimental proof has yet been found, except  for very light nuclei.

The calculations of both  of 
Refs.~\cite{Pllumbi11,Shimoyama11,Shimoyama13} are performed within the HFB+QPRA approach, and in this connection it is clearly interesting to look at the shell-model predictions. In Figs. 3 and 4, we report the calculated differential cross sections for the excitation of the ground and first excited $0^+$ states  via the ($t$,$p$) reaction on $^{134}$Sn and  the ($p$,$t$) reaction on $^{136}$Sn, respectively. 
 We see that, while for the ($t$,$p$) reaction the cross section for the excitation of the $0^+$ excited state is smaller than that 
for the excitation of the ground state (g.s.) by no more than one order of magnitude, in the ($p$,$t$) reaction the first excited $0^+$ state is very weakly populated, the cross section being four order of magnitudes smaller than that relative to the g.s.

 The latter result is strongly at variance with that obtained in 
Ref.~\cite{Pllumbi11}, where only the pick-up $^{136}$Sn($p$,$t$)$^{134}$Sn reaction was studied and the ratio of the cross sections associated to the g.s. and the $0_{2}^+$ state transitions  turned out to be not greater than 7. Our predictions are instead in agreement  with the findings of Ref.~\cite{Shimoyama11}, where both pair-addition and pair-removal transitions were considered, although most of the attention was focused on the former. 
However,  in  \cite{Shimoyama13} a microscopic analysis of the predictions reported in the study of Ref. \cite{Shimoyama11}  was performed, showing that a large number of both weakly bound and unbound states is responsible for the obtained results. 
On the other hand, our calculations‪ lead to relative cross-sections of the $L=0$ transitions, which, to a great extent, can be understood only  in terms of the transfers of $(1f_{7/2})^2$ and $(2p_{3/2})^2$ pairs.

\begin{figure}   [H]
\begin{center}
\includegraphics [scale=0.30,angle=0] {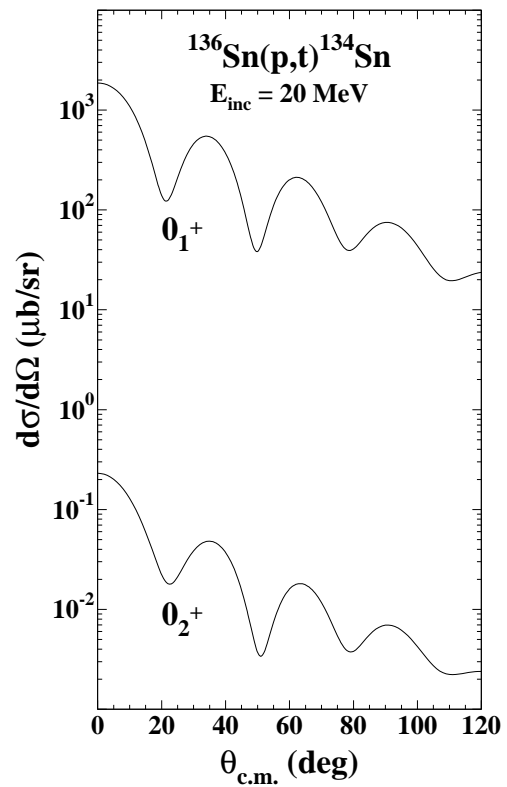}
\end{center}
\vspace{-1.2truecm}
\caption{\label{fig4} Calculated differential ($p$,$t$) cross sections  for the two lowest $0^+$ states in $^{134}$Sn at 20 MeV incident energy.    }
\end{figure}

The DWBA calculations indicate that the reaction amplitudes $f((1f_{7/2})^{2}\,;L=0;\theta)$ and $f((2p_{3/2})^{2}\,;L=0;\theta)$ of Eq. 1 are nearly independent of $\theta$, both in amplitude and in phase. Their values are in the ratio
\begin{widetext}

\begin{table}
\caption {Spectroscopic amplitudes for the population of the two lowest $0^+$ states of  $^{134}$Sn and $^{136}$Sn. The index L=0 is omitted.}
\begin{ruledtabular}
\begin{tabular}{cccc}
$nlj$&  $S((nlj)^{2}\,;\:^{136}$Sn$_{0_{1}}, ^{134}$Sn$_{0_{1}})$ & 
$S((nlj)^{2}\,; \:^{136}$Sn$_{0_{1}}, ^{134}$Sn$_{0_{2}}) $ & $S((nlj)^{2}\,; \:^{136}$Sn$_{0_{2}}, ^{134}$Sn$_{0_{1}}) $ \\
 \cline{1-4}   
$1f_{7/2}$ &    1.126 &	 -0.070 &  0.259 \\
$1f_{5/2}$ &	 0.279 &		-0.0001&  -0.082 \\
$0h_{9/2}$ &	 0.309 &		0.002 & -0.204 \\
$2p_{3/2}$ &	 0.336 &		0.030 &  -0.786 \\
$2p_{1/2}$ &	 0.183 &		0.008 &  -0.273 \\	
$0i_{13/2}$ &	 0.233 &		 -0.005 &  -0.110 \\		
\end{tabular}
\end{ruledtabular}
\label{tab2}
\end{table}
\end{widetext}

$$
\frac{f(2p_{3/2},2p_{3/2};L=0;\theta)}{f(1f_{7/2},1f_{7/2};L=0;\theta)}~\sim~1.6.
$$

The relevant spectroscopic amplitudes, $S((2p_{3/2})^{2}\,;L=0;I_1=0_1,I_2=0_{1,2})$ and $S((1f_{7/2})^{2}\,;L=0;I_1=0_1,I_2=0_{1,2})$, are presented in Table I. The sizable population of the g.s. for pickup and stripping is due to the fact that these states arise essentially from pure configurations $(1f_{7/2})^2$ and $(1f_{7/2})^4$. This may be seen in  Table II, where the wave 	functions of the two lowest $0^{+}$ states in $^{134}$Sn and $^{136}$Sn are shown by reporting, for the sake of simplicity, only the amplitudes corresponding to the two lowest-lying configurations. It is worth mentioning that the amplitudes of  the omitted configurations are all  less than 0.3. In both the transitions $^{136}$Sn$(p,t)^{134}$Sn$_{0_2}$ and $^{134}$Sn$(t,p)^{136}$Sn$_{0_2}$, there is destructive interference between $(1f_{7/2})^2$ and $(2p_{3/2})^2$ transfer, but the much larger $^{134}$Sn$(t,p)^{136}$Sn$_{0_2}$ spectroscopic amplitudes yield an appreciable $(t,p)_{0_2}$ cross-section, in spite of the destructive interference.

\begin{table} [h]
\caption {Wave functions  of  the two lowest $0^+$ states in $^{134}$Sn and $^{136}$Sn (see text for details). }
\begin{ruledtabular}
\begin{tabular}{ccccc}
& \multicolumn{2} {c} {$^{134}$Sn}  & \multicolumn{2} {c} {$^{136}$Sn} \\
 \cline{2-3}     \cline{4-5}
& ($1f_{7/2}$)$^2$ & ($2p_{3/2}$)$^2$  & ($1f_{7/2}$)$^4$& ($1f_{7/2}$)$^2$  ($2p_{3/2}$)$^2$ \\
\colrule
$0^{+}_{1}$ & 0.89 & 0.23 & 0.80   & 0.32  \\
$0^{+}_{2}$ &  0.36 & -0.83 & 0.45 & -0.72 \\		
\end{tabular} 
\end{ruledtabular}
\label{table1}
\end{table}

An even simpler view of the situation is afforded by the zero-configuration-mixing approximation, in which

\begin{eqnarray*}
|^{134}{\rm Sn,g.s.}>~&\sim&~(1f_{7/2})^2 \\
|^{134}{\rm Sn},0^+_2>~&\sim&~-(2p_{3/2})^2 \\
|^{136}{\rm Sn,g.s.}>~&\sim&~(1f_{7/2})^4 \\
|^{136}{\rm Sn},0^+_2>~&\sim&~-(1f_{7/2})^2(2p_{3/2})^2 
\end{eqnarray*}

Then both $(t,p)$ and $(p,t)$ ground state transitions would be associated with pure $(1f_{7/2})^2$ transfer, the $(t,p)$ transition to $0_2$ would be associated with pure $(2p_{3/2})^2$ transfer, whereas the $(p,t)$ transfer to $0_2$ would be completely forbidden. This is consistent with our complete calculation, in which $^{136}$Sn$(p,t)^{134}$Sn$_{0_2}$ is very weak, but $^{134}$Sn$(t,p)^{136}$Sn$_{0_2}$ is appreciable.

We can try to go a step further in this discussion  by attempting to connect the predicted structure of the wave functions to our effective interaction. We see that these wave functions are characterized by rather weak mixing, this being related to the weakness of the pairing force  with respect to the $2p_{3/2}- 1f_{7/2}$ spacing, which is about 0.9 MeV.  The two $J=0$ diagonal matrix elements of the    $(1f_{7/2})^2$ and $(2p_{3/2})^2$ configurations are both equal to about 0.7 MeV.  The non-diagonal matrix element between these two configurations  is about 0.3 MeV, which is not sufficiently large enough to generate an appreciable configuration mixing.  As mentioned in the Introduction, the reduced attractiveness of the neutron pairing force when $N>82$ has been discussed in several  previous papers  (see, for instance, \cite{Covello13}), 
and  was shown to be crucial in reproducing the observed properties in the $^{132}$Sn region beyond $N=82$. It would be clearly of great interest to see if this prediction is also confirmed  from transfer reactions  experiments.

To conclude this Section, we  show our calculated cross sections for transitions to the  yrast $2^+$, $4^+$, and $6^+$ levels in $^{136}$Sn and $^{134}$Sn. This is done in Figs. 5 and 6, where for the sake of completeness we have included the g.s. cross sections already shown in Figs. 3 and 4. In Fig. 5 we see that, according to our predictions, the three excited states should be populated with a strength not much smaller than that of the ground state.

\begin{figure}
\begin{center}
\includegraphics [scale=0.30,angle=0] {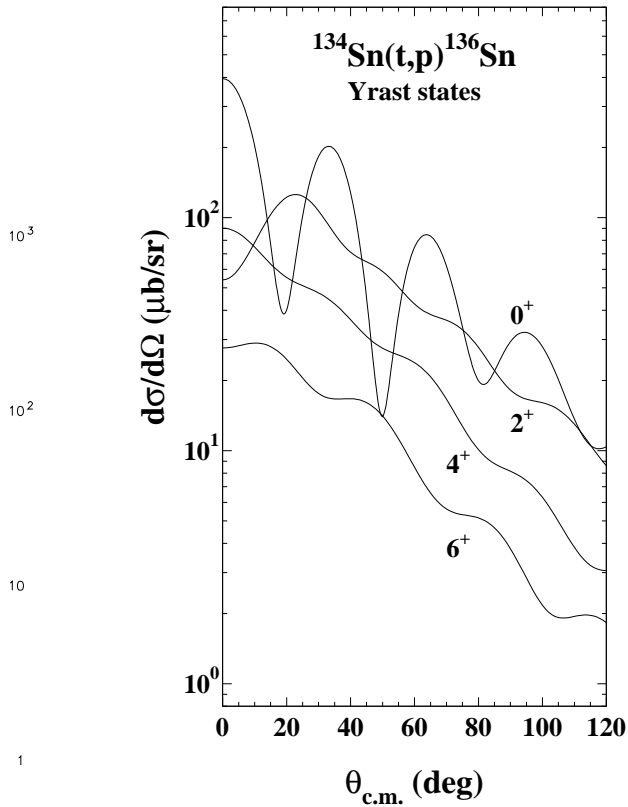}
\end{center}
\vspace{-1.2truecm}
\caption{\label{fig5}  Calculated differential ($t$,$p$) cross sections  for the yrast $0^+$, $2^+$, $4^+$, and $6^+$ levels in $^{136}$Sn at 20 MeV incident energy.   }
\end{figure}

Clearly, it be may interesting to see how these cross sections compare with those for the same reactions involving stable tin isotopes. This we do for  $^{136}$Sn($p$,$t$)$^{134}$Sn and $^{124}$Sn($p$,$t$)$^{122}$Sn, the latter reaction having been studied in our recent work \cite{Guazzoni11}. 
The two shapes for the transitions to the 0$^{+}_{1}$ state are nearly the same, as arising from the comparison between Fig. (8) of Ref. \cite{Guazzoni11} and Fig. (6) of the present paper. They decrease from the maximum at $\theta = 0^{\circ}$, and have minima at about $20^{\circ}$ and $50^{\circ}$.
However the shapes of the angular  distributions for the transions to the $2^+_1$ levels are quite different, with the $^{124}$Sn$(p,t)^{122}$Sn curve showing a rise from $\theta=0^\circ$ to maxima at about $15^\circ$ and $50^\circ$ (see Fig. 9 of \cite{Guazzoni11}), whereas the $^{136}$Sn$(p,t)^{134}$Sn curve shows no maxima, but a monotonic decrease from $\theta=0^\circ$ to $\theta=120^\circ$, on which is superposed a weak oscillation with a wavelength of about $25^\circ$.

The shapes of the angular distributions of $^{124}$Sn$(p,t)^{122}$Sn transition to the $4^+_1$ level (see Fig. (10) of \cite{Guazzoni11}) as well as that of  $^{136}$Sn$(p,t)^{134}$Sn transition to the same level show gentle oscillations, but they are approximately $180^\circ$ out of phase with each other. The $^{124}$Sn$(p,t)^{122}$Sn angular distribution to the $6^+_1$ level rises steadily to a maximum at about $50^\circ$ (see Fig. (11) of \cite{Guazzoni11}),  whereas the $^{136}$Sn$(p,t)^{134}$Sn angular distribution to this level is similar  to the $^{124}$Sn$(p,t)^{122}$Sn $2^+_1$ level, falling continuously with weak oscillations.

The basic reason for such differences between the two reactions is the large $Q$-value difference.  Due to the weak binding of the outer neutrons in $^{136}$Sn, the $Q$-value of the $^{136}$Sn($p$,$t$)$^{134}$Sn  reaction is positive, so the outgoing triton has more momentum than in the case of  $^{124}$Sn($p$,$t$)$^{122}$Sn, which has the usual negative ($p$,$t$) $Q$-value. 
The large differences between the $^{136}$Sn($p$,$t$)$^{134}$Sn  and $^{124}$Sn($p$,$t$)$^{122}$Sn   momentum transfers account for the differences in the shapes of the ($p$,$t$) angular distributions. We have found similar $Q$-value dependence in the comparison of angular distribution shapes of the $^{A}$Sn($t$,$p$)$^{A+2}$Sn stripping reactions to the yrast levels  ($2^+$,$4^+$,$6^+$). 

\begin{figure} [H]
\begin{center}
\includegraphics [scale=0.30,angle=0] {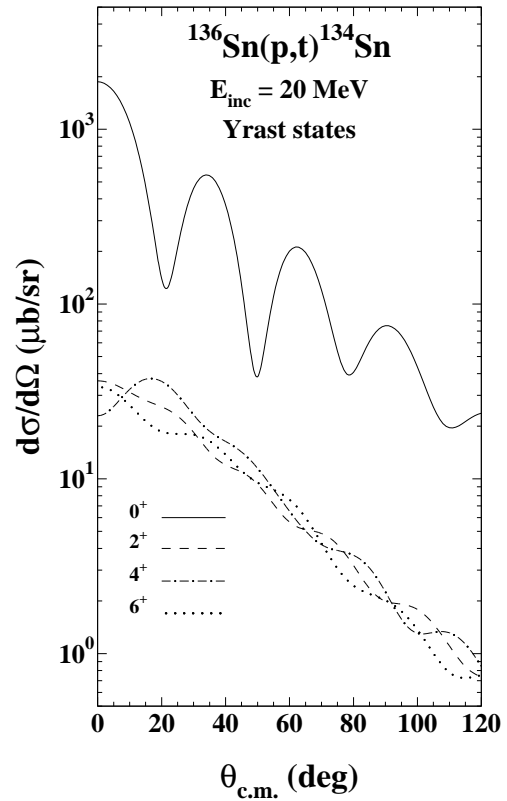}
\end{center}
\vspace{-1.2truecm}
\caption{\label{fig6}  Calculated differential ($p$,$t$) cross sections  for the yrast $0^+$, $2^+$, $4^+$, and $6^+$ levels in $^{134}$Sn at 20 MeV incident energy.   }
\end{figure}
\section{Summary}

We have presented here the results of a study of two-neutron transfer reactions on the first two Sn isotopes beyond $N=82$, $^{134}$Sn and $^{136}$Sn, which are likely to be studied experimentally in the near future, thanks to the availability of RIBs of sufficiently high intensity. In this study we have used, for the states of both nuclei, wave functions obtained from realistic shell-model calculations without use of any adjustable parameters.
It is worth emphasizing that these calculations reproduced, with great accuracy,  the energies of the few available observed levels.
We have calculated the differential cross sections for transitions to the two lowest 
$0^+$ states as well as to  the yrast $2^+$, $4^+$ and $6^+$ levels in both nuclei. 
The results obtained confirm the idea that the reaction $^{134}$Sn$(t,p)^{136}$Sn is of utmost interest to shed light on the structure of neutron-rich Sn isotopes adjacent to doubly-magic $^{132}$Sn. We hope our predictions may stimulate efforts in this direction.

\end{document}